\documentclass[12pt]{article}
\usepackage{latexsym}
\usepackage{epsfig}
\usepackage{amsmath}
\usepackage{amssymb}
\usepackage{epsf}
\usepackage{epsfig}
\usepackage{cite}
\newcommand{\lsim}{
\mathrel{\hbox{\rlap{\hbox{\lower4pt\hbox{$\sim$}}}\hbox{$<$}}}}

\newcommand{\gsim}{
\mathrel{\hbox{\rlap{\hbox{\lower4pt\hbox{$\sim$}}}\hbox{$>$}}}}

\def\D0{D\O }

\def\Adel{\mathcal{A}^f_{\Delta\Gamma}}

\begin{document}
\begin{titlepage}
\vspace*{-0.0truecm}

\begin{flushright}
Nikhef-2011-026
\end{flushright}

\vspace*{1.8truecm}

\begin{center}
\boldmath
{\Large{\bf Effective Lifetimes of $B_s$ Decays and their\\ 
Constraints on the $B_s^0$--$\bar B_s^0$ Mixing Parameters}}
\unboldmath
\end{center}

\vspace{0.9truecm}

\begin{center}
{\bf Robert Fleischer  \,and\, Robert Knegjens}

\vspace{0.5truecm}

{\sl Nikhef, Science Park 105, NL-1098 XG Amsterdam, The Netherlands}
 
\end{center}

\vspace{1.6cm}
\begin{abstract}
\vspace{0.2cm}\noindent
Measurements of the effective lifetimes of $B_s$-meson decays, which only require untagged rate
analyses, allow us to probe the width difference $\Delta\Gamma_s$ and the CP-violating phase 
$\phi_s$ of $B^0_s$--$\bar B^0_s$ mixing. We point out that the dependence of the effective 
lifetime on non-linear terms in $\Delta\Gamma_s$ allows for a determination of $\phi_s$ and 
$\Delta\Gamma_s$ given a pair of $B_s$ decays into CP-even and CP-odd final states. 
Using recent lifetime measurements of $B^0_s\to K^+K^-$ and $B^0_s\to J/\psi f_0(980)$ 
decays, we illustrate this method and show how it complements the constraints in the 
$\phi_s$--$\Delta\Gamma_s$ plane from other observables. 
\end{abstract}

\vspace*{0.5truecm}
\vfill
\noindent
September 2011
\vspace*{0.5truecm}

\end{titlepage}

\thispagestyle{empty}
\vbox{}
\newpage

\setcounter{page}{1}

\section{Introduction}\label{sec:intro}
A promising avenue for New Physics (NP) to enter the observables of $B_s$-meson decays
is given by $B^0_s$--$\bar B^0_s$ mixing. In the Standard Model (SM), the phenomenon of mixing
originates from box topologies and is strongly suppressed. In the presence of NP,  
new particles could give rise to additional box topologies or even contribute at the tree level. 
Should these NP contributions also involve new CP-violating phases, the 
$B^0_s$--$\bar B^0_s$ mixing phase $\phi_s$ could differ sizably from the tiny SM 
value of $-2.1^\circ$ (see, for instance, Refs.~\cite{Bmix,DFN} and references therein). 

A key channel for addressing this exciting possibility is $B^0_s\to J/\psi \phi$. 
A characteristic feature of this channel is that its final state contains two vector mesons 
and thereby requires a time-dependent angular analysis of the 
$J/\psi \to \mu^+\mu^-$ and $\phi\to K^+K^-$ decay products \cite{DDF,DFN}. 
Over the last couple of years, measurements at the Tevatron of 
CP-violating asymmetries in ``tagged" analyses (distinguishing between initially 
present $B^0_s$ or $\bar B^0_s$  mesons) of the $B^0_s\to J/\psi \phi$ channel indicate 
possible NP effects in $B^0_s$--$\bar B^0_s$ mixing \cite{CDF-phis,D0-phis,TEV-LP}. 
These results are complemented by the measurement of the anomalous like-sign dimuon 
charge asymmetry at \D0, which was found to differ by $3.9\,\sigma$ from the SM prediction 
\cite{di-muon}. This summer, the LHCb collaboration has also joined the arena, reporting, 
however, results that disfavour large NP effects \cite{LHCb-LP}. The above measurements, 
which we will discuss in more detail below, are typically shown in the 
$\phi_s$--$\Delta\Gamma_s$ plane, where $\Delta\Gamma_s$ is the width difference 
between the mass eigenstates of the $B_s$-meson system. 

In this paper, we point out a new method for determining further constraints in the 
$\phi_s$--$\Delta\Gamma_s$ plane using measurements of the effective lifetimes of 
$B_s$ decays. In particular, we show that the information provided by the 
lifetimes of a pair of decays into CP-even and CP-odd final states is sufficient to determine 
$\phi_s$ and $\Delta\Gamma_s$. The advantage of this strategy is that it only requires an 
``untagged" analysis, i.e.\ it is not necessary to distinguish between initially present $B^0_s$ or 
$\bar B^0_s$ mesons, which is experimentally advantageous. 

Specifically, we will consider the $B^0_s\to K^+K^-$ \cite{RF-BsKK,FK} and 
$B^0_s\to J/\psi f_0(980)$ \cite{SZ,FKR} decays, which have final states with the CP eigenvalues 
$+1$ and $-1$, respectively. From here on we shall abbreviate the latter decay as 
$B^0_s\to J/\psi f_0$. First measurements of the effective lifetimes of these channels 
are already available from the CDF  and LHCb collaborations 
\cite{CDF-KK-lifetime,Aaltonen:2011nk,LHCb-KK-lifetime}. For the theoretical interpretation of 
these results we also need to address hadronic uncertainties. A closer look will reveal that these 
decays are well suited in this respect. We will illustrate our method with the most recent 
data and shall compare the resulting constraints in the $\phi_s$--$\Delta\Gamma_s$ plane with 
those from the alternative measurements listed above. 

The outline is as follows: in Section~\ref{sec:effective}, we discuss the general 
formalism to calculate effective lifetimes and show in Section~\ref{sec:cubic} how the 
corresponding measurements can be converted into contours in the $\phi_s$--$\Delta\Gamma_s$
plane. In Section~\ref{sec:control}, we turn to the 
hadronic uncertainties affecting this analysis and their control through experimental data. 
The constraints on the $B^0_s$--$\bar B^0_s$ mixing parameters arising from the current data 
for the effective lifetimes of the $B^0_s\to K^+K^-$ and $B^0_s\to J/\psi f_0$ channels are 
explored in Section~\ref{sec:constraints}, where we also illustrate the impact of future
lifetime measurements with errors at the $1\%$ level. In Section~\ref{sec:future}, we
give a collection of additional $B_s$ decays that can be added to this analysis in the future. 
Finally, we summarize our conclusions in Section~\ref{sec:concl}.

\section{General Formalism}\label{sec:effective}
We will consider a $B_s\to f$ transition with a final state $f$ into which both a $B^0_s$ and a 
$\bar B^0_s$ meson can decay. The corresponding untagged rate can then be written as 
follows \cite{DFN}:
\begin{align}\label{eqn:untagged}
	\langle \Gamma(B_s(t)\to f)\rangle 
	&\equiv \Gamma(B^0_s(t)\to f)+ \Gamma(\bar B^0_s(t)\to f) \notag\\
	&=  R_{{\rm H}}^f\, e^{-\Gamma_{\rm H}^{(s)} t} 
	+ R_{{\rm L}}^f\, e^{-\Gamma_{\rm L}^{(s)} t},
\end{align}
where L and H denote the light and heavy $B_s$ mass eigenstates, respectively. Using
\begin{equation}\label{eqn:DGdefn}
\Gamma_s \equiv \frac{\Gamma_{\rm L}^{(s)} + \Gamma_{\rm H}^{(s)}}{2} = \tau_{B_s}^{-1},\quad 
\Delta \Gamma_s \equiv \Gamma_{\rm L}^{(s)} - \Gamma_{\rm H}^{(s)},
\end{equation}
we can straightforwardly write (\ref{eqn:untagged}) as
\begin{align}\label{eqn:adelIntro}
	\langle \Gamma(B_s(t)\to f)\rangle
	\propto\ e^{-\Gamma_st}\left[ \cosh\left(\frac{\Delta\Gamma_s t}{2}\right)
	 + {\cal A}_{\Delta\Gamma}^f\,\sinh\left(\frac{\Delta\Gamma_s t}
	{2}\right)\right]
\end{align}
with
\begin{equation}
	{\cal A}_{\Delta\Gamma}^f
	\equiv \frac{R_{{\rm H}}^f - R_{{\rm L}}^f}
			{R_{{\rm H}}^f + R_{{\rm L}}^f}.
\end{equation}

We define the effective lifetime of the decay $B^0_s\to f$ as the time expectation value 
of the untagged rate \cite{FK},
\begin{equation}\label{eqn:efftime}
  \tau_{f} 
  \equiv \frac{\int^\infty_0 t\ \langle \Gamma(B_s(t)\to f)\rangle\ dt}
  {\int^\infty_0 \langle \Gamma(B_s(t)\to f)\rangle\ dt} =
  \frac{ R_{{\rm L}}^f/\Gamma_{\rm L}^{(s)2} + R_{{\rm H}}^f/\Gamma_{\rm H}^{(s)2} }
  {R_{{\rm L}}^f/\Gamma_{\rm L}^{(s)}+ R_{{\rm H}}^f/\Gamma_{\rm H}^{(s)}},
  \end{equation}
which is equivalent to the lifetime that results from fitting the two 
exponentials in \eqref{eqn:untagged} to a single exponential \cite{Hartkorn:1999ga}.
By making the usual definition 
\begin{equation}
y_s\equiv \frac{\Delta\Gamma_s}{2\Gamma_s}, 
\end{equation}
we can express the effective lifetime as
\begin{align}
	\frac{\tau_f}{\tau_{B_s}}
	&= \frac{1}{1-y_s^2} \left( \frac{1+2{\cal A}^f_{\Delta\Gamma}\, y_s + y_s^2}
	{1+{\cal A}^f_{\Delta\Gamma}\, y_s}\right)\nonumber\\
	&= 1 + \Adel\,y_s + \left[2- (\Adel)^2\right]y_s^2 + {\cal O}(y_s^3), \label{eqn:tauExact}
\end{align}
where we have also given the expansion in powers of $y_s$ up to cubic corrections. 

We proceed to consider the case where $f$ is a CP eigenstate with eigenvalue $\eta_f$.
In the SM, the decay amplitude can be written, without loss of generality 
(using the unitarity of the Cabibbo--Kobayashi--Maskawa (CKM) matrix), as
\begin{equation}
A(B^0_s\to f)=A_1^fe^{i\delta_1^f}e^{i\varphi_1^f}+A_2^fe^{i\delta_2^f}e^{i\varphi_2^f},
\end{equation}
where the $A_{1,2}^f$ are real and the $\delta_{1,2}^f$ and $\varphi_{1,2}^f$ are 
CP-conserving strong and CP-violating weak phases, respectively. Using 
the standard $B_s^0$--${\bar B}_s^0$ mixing formalism \cite{RF-habil}, we have
\begin{equation}\label{eqn:Adel}
	{\cal A}_{\Delta\Gamma}^f  = \frac{2\,{\rm Re}\,\xi_f^{(s)}}{1+\bigl|\xi_f^{(s)}\bigr|^2},
\end{equation}
where
\begin{equation}\label{eqn:xiMix}
	\xi_f^{(s)} = - \eta_f e^{-i\phi_s} \left[\frac{e^{-i\varphi_1^f}+
	h_fe^{i\delta_f}e^{-i\varphi_2^f}}{e^{i\varphi_1^f}+h_fe^{i\delta_f}e^{i\varphi_2^f}}\right].
\end{equation}
Here we have introduced the abbreviation
\begin{equation}\label{h-def}
h_fe^{i\delta_f}\equiv\frac{A_2^f}{A_1^f}e^{i(\delta_2^f-\delta_1^f)},
\end{equation}
and $\phi_s$ denotes the $B_s^0$--${\bar B}_s^0$ mixing phase, which is given by
\begin{equation}\label{phis}
\phi_s\equiv\phi_s^{\rm SM}+\phi_s^{\rm NP},
\end{equation}
where $\phi_s^{\rm SM}$ and $\phi_s^{\rm NP}$ are the SM and NP pieces, respectively.
It is convenient for the following discussion to introduce the direct CP asymmetry 
of the $B_s \to f$ decay \cite{RF-habil}:
\begin{equation}
	C_f \equiv \frac{1-\left|\xi_f\right|^2}{1+\left|\xi_f\right|^2}=
	\frac{2\,h_f\sin\delta_f\sin(\varphi_1^f-\varphi_2^f)}{N_f},
\end{equation}
where
\begin{equation}
N_f\equiv 1+ 2 h_f \cos\delta_f\cos(\varphi_1^f-\varphi_2^f)+h_f^2.
\end{equation}
Subsequently, we may write 
\begin{equation}\label{xi-simpl}
\frac{2\,\xi_f^{(s)}}{1+\bigl|\xi_f^{(s)}\bigr|^2}=-\eta_f\sqrt{1-C_f^2}\,e^{-i(\phi_s+\Delta\phi_f)}.
\end{equation}
Here $\Delta\phi_f$ is a hadronic phase shift, which is given by
\begin{equation}\label{s-expr}
\sin\Delta\phi_f=\frac{\sin2\varphi_1^f+2h_f\cos\delta_f\sin(\varphi_1^f+\varphi_2^f)+
h_f^2\sin2\varphi_2^f}{N_f\sqrt{1-C_f^2}}
\end{equation}
\begin{equation}\label{c-expr}
\cos\Delta\phi_f=\frac{\cos2\varphi_1^f+2h_f\cos\delta_f\cos(\varphi_1^f+\varphi_2^f)+
h_f^2\cos2\varphi_2^f}{N_f\sqrt{1-C_f^2}},
\end{equation}
yielding
\begin{equation}
\tan\Delta\phi_f=\frac{\sin2\varphi_1^f+2h_f\cos\delta_f\sin(\varphi_1^f+\varphi_2^f)+
h_f^2\sin2\varphi_2^f}{\cos2\varphi_1^f+2h_f\cos\delta_f\cos(\varphi_1^f+\varphi_2^f)+
h_f^2\cos2\varphi^f_2}.
\end{equation}
The twofold ambiguity for $\Delta\phi_f$ arising from the latter expression can be resolved
using sign information from $\sin\Delta\phi_f$ or $\cos\Delta\phi_f$. These
expressions generalize those given in Refs.~\cite{FKR,FFM}.  

Using (\ref{eqn:Adel}) and (\ref{xi-simpl}), we thus obtain 
\begin{equation}\label{eqn:AdelCos}
	{\cal A}_{\Delta\Gamma}^f = - \eta_f \sqrt{1- C_f^2}\, \cos (\phi_s + \Delta\phi_{f}).
\end{equation}
As we will see in Section~\ref{sec:control}, there are fortunate $B_s$ decays into CP
eigenstates where the hadronic parameter $h_fe^{i\delta_f}$ and the resulting phase shift 
$\Delta\phi_f$ can be controlled through experimental data. For these decays, we can hence 
use the corresponding lifetime measurements to constrain $y_s$ (or $\Delta\Gamma_s$) 
with respect to $\phi_s$.

\boldmath
\section{Lifetime Contours in the $\phi_s$--$\Delta\Gamma_s$ Plane}\label{sec:cubic}
\unboldmath
Let us now have a closer look at  \eqref{eqn:tauExact}, which we can write as the 
following cubic equation for the real parameter $y_s$:
\begin{equation}\label{cubic-1}
y_s^3+a_2y_s^2+a_1y_s+a_0=0,
\end{equation}
where
\begin{align}
	a_0 \equiv\ \frac{\tau_{B_s} - \tau_f}{\tau_f {\cal A}^f_{\Delta\Gamma}},\quad
	a_1 \equiv\  \frac{2\,\tau_{B_s} - \tau_f}{\tau_f },\quad
	a_2 \equiv\ \frac{\tau_{B_s} + \tau_f}{\tau_f {\cal A}^f_{\Delta\Gamma}}.
\end{align}
In order to solve this cubic equation, it is useful to rewrite it in the ``reduced" form
\begin{equation}
\left( y_s+\frac{a_2}{3} \right)^3+ P \left( y_s+\frac{a_2}{3}  \right) + Q =0
\end{equation}
with
\begin{equation}
P\equiv a_1-\frac{a_2^2}{3}, \quad Q\equiv \frac{2a_2^3}{27}-\frac{a_2a_1}{3}+a_0.
\end{equation}
Applying Cardano's formula then yields the solutions
\begin{align}\label{exact-sol}
	y_s = -\frac{a_2}{3} &+ e^{i\omega}\, \sqrt[3]{R+\sqrt{D}} 
	+ e^{-i\omega}\, \sqrt[3]{R-\sqrt{D}}
\end{align}
with $\omega \in \{0,2\pi/3,-2\pi/3\}$, where
\begin{equation}
 	R \equiv-\frac{Q}{2}=\ \frac{1}{54}\left(9\,a_1\, a_2 - 27\, a_0 - 2\, a_2^3 \right)
\end{equation}
\begin{equation}		
	D \equiv\left(\frac{P}{3}\right)^3+\left(\frac{Q}{2}\right)^2
	= \frac{1}{108}\left(27\, a_0^2 - 18\, a_0\, a_1\, a_2 + 4\, a_0\, a_2^3
	+ 4\, a_1^3 -a_1^2 a_2^2 \right).
\end{equation}

For ${\cal A}^f_{\Delta\Gamma}=0$, this solution is not valid as \eqref{eqn:tauExact} is then a 
quadratic equation in $y_s$.
Furthermore, the above expressions may prove cumbersome to use in practice.
A convenient approximate solution is obtained by solving the expansion in (\ref{eqn:tauExact}) 
up to quadratic order in $y_s$: 
\begin{equation}
y_s\approx-\frac{1}{2}\left[\frac{{\cal A}^f_{\Delta\Gamma}}{2-({\cal A}^f_{\Delta\Gamma})^2}\right]
\pm\frac{1}{2}\sqrt{\left[\frac{{\cal A}^f_{\Delta\Gamma}}{2-({\cal A}^f_{\Delta\Gamma})^2}\right]^2
+\frac{4}{\tau_{B_s}}\left[\frac{\tau_f - \tau_{B_s}}{2-({\cal A}^f_{\Delta\Gamma})^2}\right]}.
\end{equation}
This quadratic solution is in excellent agreement with the corresponding branches of the 
exact solution (\ref{exact-sol}) for the numerical analyses discussed below.

\begin{figure}[tbp] 
   \centering
   \includegraphics[width=7.9truecm]{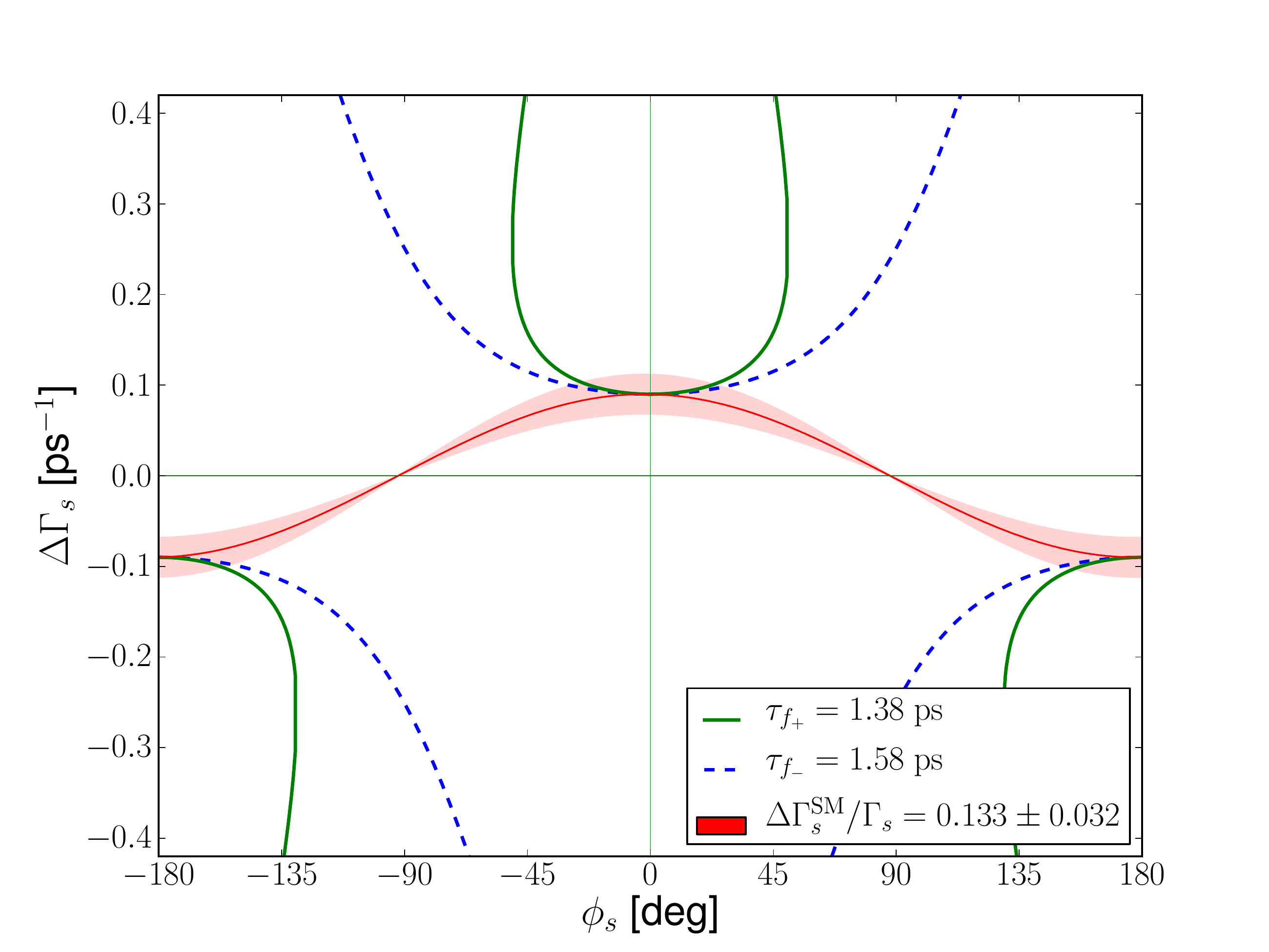} 
   \caption{Illustration of the lifetimes that are compatible with the SM value of 
   $\Delta\Gamma_s/\Gamma_s$ given in \eqref{eqn:DGGSM} for CP-even and 
   CP-odd final states $f_+$ and $f_-$, respectively. The decay amplitudes are assumed 
   to have no CP-violating phases. We also show the constraint from the theoretical
   value of $\Delta\Gamma_s^{\rm SM}/\Gamma_s$ given in (\ref{eqn:DGGSM}), as 
   discussed in the text.}\label{fig:CPillust}
\end{figure}

For illustration we consider two $B_s$ decays to CP eigenstates, $B_s\to f_+$ and $B_s\to f_-$, 
with positive and negative CP eigenvalues, respectively. Further, we assume 
\begin{equation}
h_{f_\pm}=0, \quad \varphi_1^{f_\pm}=0 
\end{equation}
for these decays, yielding $C_{f_\pm}=0$ and $\Delta\phi_{f_\pm} = 0$. In Fig.~\ref{fig:CPillust},
we show the lifetime constraints that are compatible with the theoretical SM calculation of 
$\Delta\Gamma_s$ \cite{NL},
\begin{equation}\label{eqn:DGGSM}
	 \frac{\Delta\Gamma^{\rm SM}_s}{\Gamma_s} = 2\, y_{s}^{\rm SM} =  0.133\pm 0.032,
\end{equation}
and the SM value of the $B^0_s$--$\bar B^0_s$ mixing phase, which is given as
follows \cite{Charles:2011va}:
\begin{equation}
	\phi_s^{\rm SM} \equiv -2\beta_s = -(2.08 \pm 0.09)^\circ.
\end{equation}
Throughout this paper, we shall use \cite{HFAG}
\begin{equation}\label{eqn:tauBs}
	\tau_{B_s} = \left(1.477^{+0.021}_{-0.022}\right){\rm ps}
\end{equation}
for the $B_s$ lifetime introduced in \eqref{eqn:DGdefn}, resulting in the SM effective 
lifetimes $\tau_{f_+} = 1.38$\,ps and $\tau_{f_-} = 1.58$\,ps. The difference in behaviour for 
CP-odd and CP-even eigenstates is due to the non-linear dependence on $y_s$ in 
\eqref{eqn:tauExact}. Said differently, if \eqref{eqn:tauExact} is expanded and only terms 
up to linear order in $y_s$ are kept the two curves in Fig.~\ref{fig:CPillust} would overlap.

In Fig.~\ref{fig:CPillust}, we have included another constraint, which is related to the
theoretical value in (\ref{eqn:DGGSM}) as follows: if we assume that NP can only affect 
$\Delta\Gamma_s$ through $B^0_s$--$\bar B^0_s$ mixing, which is a very plausible 
assumption, we have \cite{Grossman}
\begin{equation}\label{ys}
y_s=\frac{\Delta\Gamma_s^{\rm SM}\cos\tilde\phi_s}{2\Gamma_s}=
y_s^{\rm SM}\cos\tilde\phi_s,
\end{equation}
where
\begin{equation}
\tilde\phi_s\equiv\tilde\phi_s^{\rm SM}+\phi_s^{\rm NP}.
\end{equation}
Here $\phi_s^{\rm NP}$ is the NP $B^0_s$--$\bar B^0_s$ mixing phase,
which also enters $\phi_s$ defined in (\ref{phis}) on which $\Adel$ depends, 
whereas the SM piece takes the following value \cite{NL}:
\begin{equation}\label{tildephis}
\tilde\phi_s^{\rm SM}=(0.22\pm0.06)^\circ.
\end{equation}

The formalism developed above is also valid for non-CP eigenstates provided the final state is 
accessible to both $B^0_s$ and $\bar B^0_s$ so that mixing is possible.
Examples of such states are $B_s \to D_s^\pm K^{(*)\mp}$.
For these decays the CP eigenvalue $\eta_f$ in \eqref{eqn:AdelCos} should be replaced by 
$(-1)^L$, where $L$ denotes the relative orbital angular momentum of the decay 
products \cite{RF-BsDsK}.

\section{Hadronic Corrections and Their Control}\label{sec:control}
Examples of effective lifetimes that have been measured for $B_s$ decays to CP-even and 
CP-odd final states are $B^0_s\to K^+ K^-$ and $B^0_s \to J/\psi f_0$, respectively.
Unlike our hypothetical examples from the previous section, however, the decay amplitudes 
of these decays are not devoid of weak phases, and can be written in the SM as follows 
\cite{RF-BsKK,FK,FKR}:
\begin{equation}
A(B^0_s\to K^+K^-)=\lambda\,{\cal C}\left[e^{i\gamma}+\frac{1}{\epsilon}de^{i\theta}\right]
\end{equation}
\begin{equation}
A(B^0_s\to J/\psi f_0)=\left(1-\frac{\lambda^2}{2}\right) {\cal A}\left[1+
\epsilon b e^{i\vartheta}e^{i\gamma} \right]. 
\end{equation}
Here $\lambda\equiv|V_{us}|=0.2252 \pm 0.0009$ is the Wolfenstein parameter of the 
CKM matrix \cite{PDG}, 
\begin{equation}
\epsilon\equiv\frac{\lambda^2}{1-\lambda^2}=0.0534  \pm 0.0005
\end{equation}
and $\gamma$ is the usual angle of the unitarity triangle whereas ${\cal C}$, $de^{i\theta}$
and ${\cal A}$, $b e^{i\theta}$ are hadronic, CP-conserving parameters. 

Consequently, the parameters introduced in \eqref{eqn:Adel} and (\ref{h-def}) take the forms
\begin{equation}
h_{K^+K^-}=d/\epsilon, \quad \delta_{K^+K^-}=\theta, \quad \varphi_1^{K^+K^-}=\gamma, 
\quad  \varphi_2^{K^+K^-}=0
\end{equation}
and
\begin{equation}
h_{J/\psi f_0}=\epsilon b, \quad \delta_{J/\psi f_0}=\vartheta, \quad \varphi_1^{J/\psi f_0}=0, 
\quad  \varphi_2^{J/\psi f_0}=\gamma,
\end{equation}
so that the hadronic phase shifts can be obtained from 
\begin{equation}\label{eqn:delKK}
\tan\Delta\phi_{K^+K^-}=2\epsilon\left[\frac{d\cos\theta + \epsilon\cos\gamma}
{d^2 + 2\epsilon\, d \cos\theta\cos\gamma + \epsilon^2 \cos 2\gamma}\right]\sin\gamma,
\end{equation}
\begin{equation}\label{tDelphi}
\tan\Delta\phi_{J/\psi f_0}=2\epsilon b \left[\frac{\cos\vartheta +\epsilon b
\cos\gamma}{1+ 2 \epsilon b\cos\vartheta\cos\gamma+\epsilon^2b^2\cos2\gamma}
\right]\sin\gamma.
\end{equation}
We observe that these phases are proportional to the tiny $\epsilon$ 
parameter, i.e.\ are doubly Cabibbo-suppressed. Consequently, the ${\cal A}_{\Delta\Gamma}^f $
observable given in (\ref{eqn:AdelCos}) is robust with respect to hadronic uncertainties
for the decays at hand.

In the case of the $B^0_s\to K^+K^-$ channel, we can use the $U$-spin symmetry of 
strong interactions to relate it to the $B^0_d\to \pi^+\pi^-$ decay, which thereby 
allows us to determine $\gamma$ as well as $d$ and $\theta$ \cite{RF-BsKK}. 
The current status of an analysis along
these lines, using also the direct CP violation in $B_d\to\pi^\mp K^\pm$, is given 
by \cite{FK}:
\begin{equation}\label{eqn:gamma}
	\gamma = (68 \pm 7)^\circ,  \quad d=0.50^{+0.12}_{-0.11}, \quad \theta=(154^{+11}_{-14})^\circ,
\end{equation}
where the errors include the uncertainties of the relevant input quantities and estimates of 
$U$-spin-breaking corrections. The $\gamma$ result is in excellent agreement with the 
current fits of the unitarity triangle \cite{CKMfitter,UTfit}, thereby excluding large CP-violating 
NP contributions to the $B^0_s\to K^+K^-$ decay amplitude. Using these numbers 
in \eqref{eqn:delKK}, we find
\begin{equation}\label{eqn:DelPhif0}
\Delta\phi_{K^+K^-} = -\left(10.5{}^{+0.3}_{-0.5}\bigl|_{\gamma}
{}^{+2.9}_{-2.1}\bigl|_{d}{}^{+0.9}_{-1.7}\bigl|_{\theta}
\right)^\circ = -\left(10.5{}^{+3.1}_{-2.8}\right)^\circ, 
\end{equation}
where we have added the errors in quadrature. Similarly, we also find
$C_{K^+ K^-} = 0.09^{+0.05}_{-0.04}$.

Unfortunately, as discussed in Ref.~\cite{FKR}, it is much more involved to control the
hadronic effects in the $B^0_s\to J/\psi f_0$ decay through experimental data, 
and the potential control channel $B^0_d\to J/\psi f_0$ has not yet been observed. 
On the other hand, contrary to $B^0_s\to K^+ K^-$, the denominator of (\ref{tDelphi}) 
is equal to one at leading order in $\epsilon$. Following Ref.~\cite{FKR}, we use the 
conservative range $0\leq b \leq 0.5$ and leave $\vartheta$ unconstrained. Using
moreover the value for $\gamma$ in \eqref{eqn:gamma}, we find
\begin{equation}\label{eqn:DelPhiKK}
\Delta\phi_{J/\psi f_0} \in \left[ -2.9^\circ,2.8^\circ \right]
\end{equation}
and $\left|C_{J/\psi f_0}\right| \lesssim 0.05$, which has, just like 
$C_{K^+ K^-}$, a negligible impact on (\ref{eqn:AdelCos}). 

It is remarkable that the hadronic phase shifts $\Delta\phi_{K^+ K^-}$ and $\Delta\phi_{J/\psi f_0}$ 
turn out to be so robust with respect to the hadronic effects and the weak phase $\gamma$,
suffering from uncertainties of only $\sim 3^\circ$. Future data should allow us to determine 
them with even higher precision.

\begin{figure}[tbp] 
   \centering
      \begin{tabular}{cc}
   	  \includegraphics[width=7.9truecm]{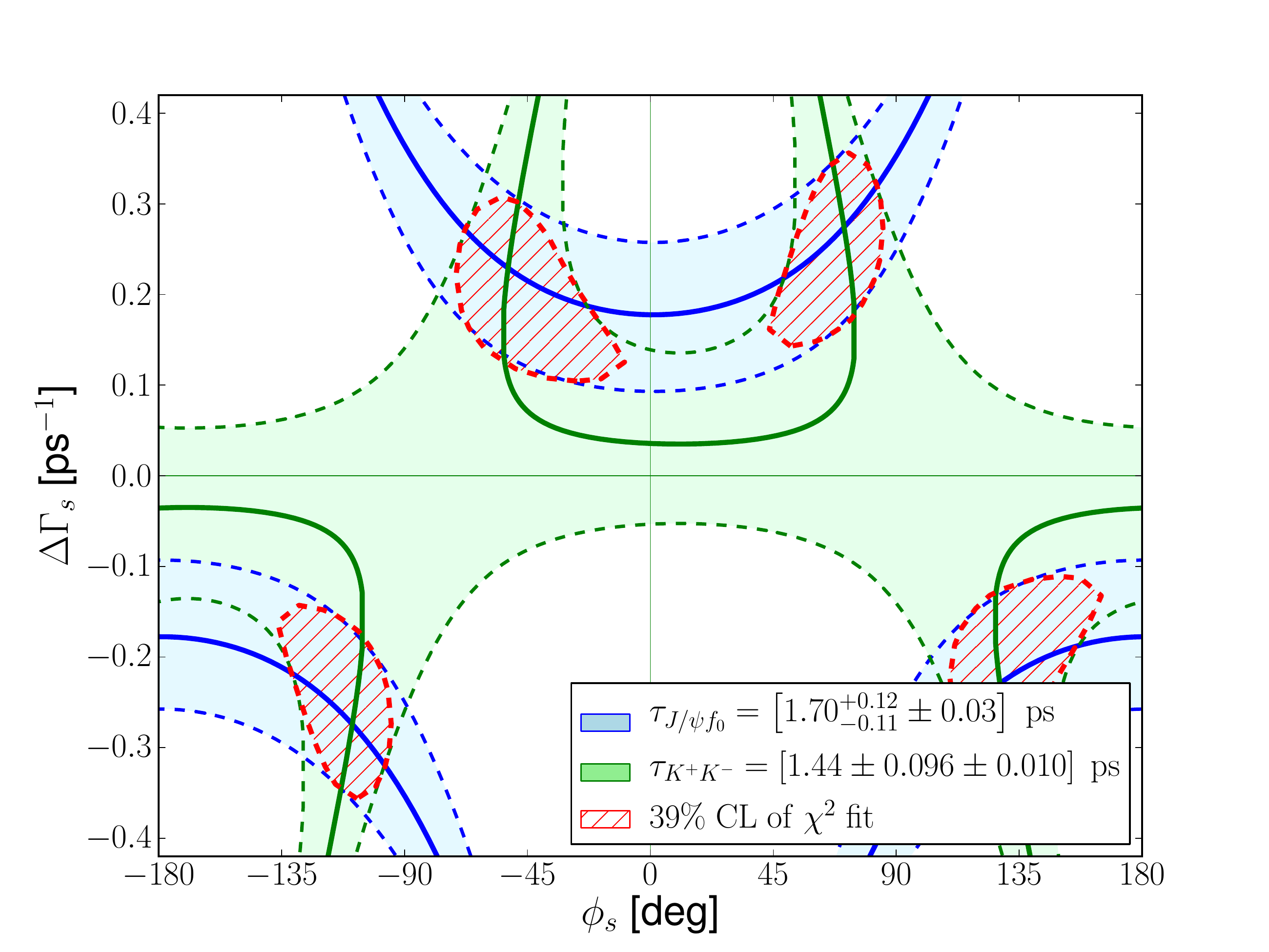} &
      \includegraphics[width=7.9truecm]{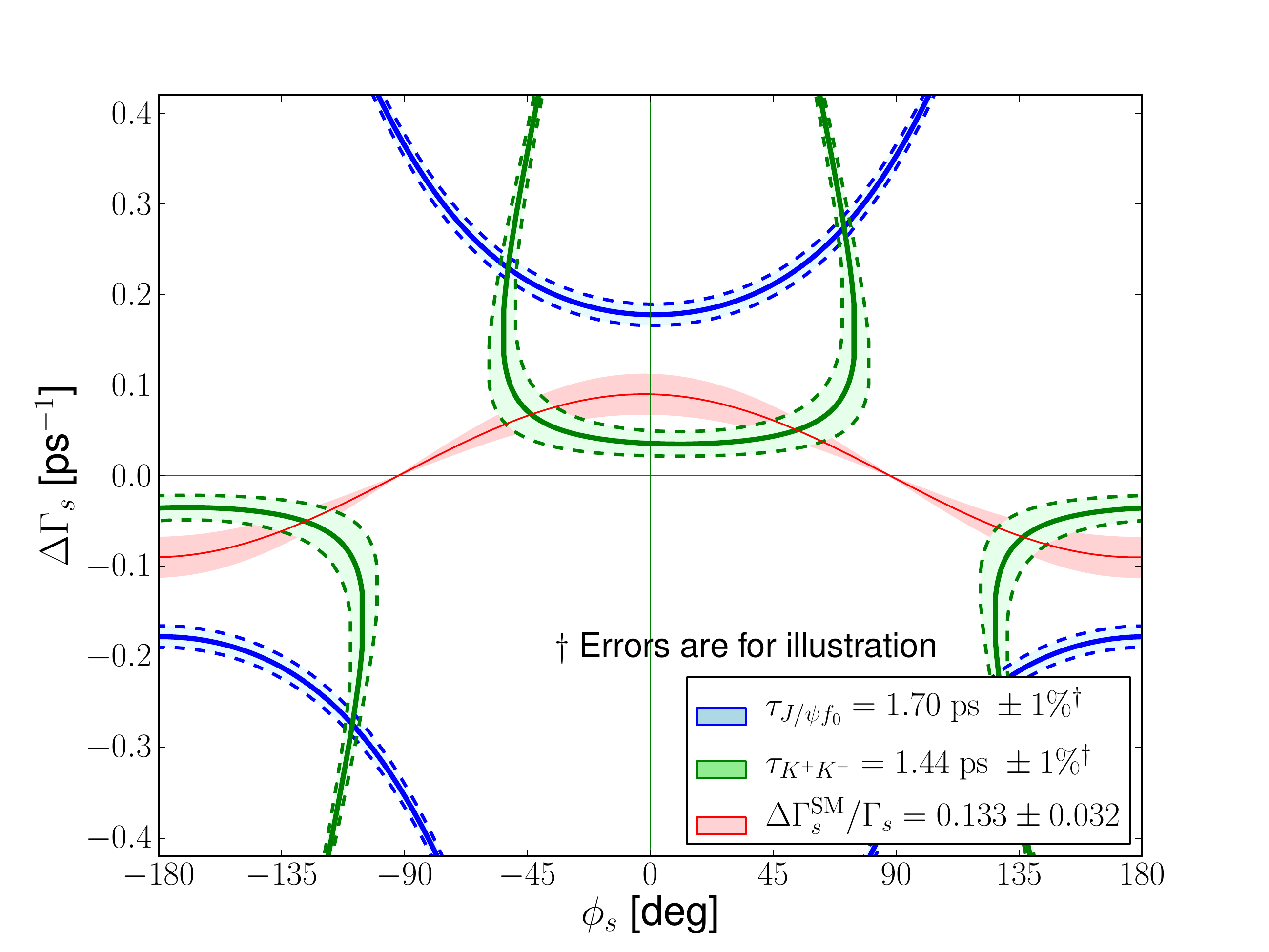}
   \end{tabular}
     \caption{The measurements of the effective $B^0_s\to K^+ K^-$ 
   and $B^0_s\to J/\psi f_0$ lifetimes projected onto the $\phi_s$--$\Delta\Gamma_s$ plane. 
   {\it Left panel:} analysis of the current data, where the shaded bands give the 
   1\,$\sigma$ uncertainties of the lifetimes; the 39\% confidence regions originating from 
   a $\chi^2$ fit are also shown. {\it Right panel}: illustration of how the situation improves 
   for unchanged central values if the uncertainties were improved to 1\% accuracy, including
   also the constraint from the theoretical value of 
   $\Delta\Gamma_s^{\rm SM}/\Gamma_s$.}\label{fig:DGphiS}
\end{figure}

\boldmath
\section{Constraints from Current and Future Data}\label{sec:constraints}
\unboldmath
In the previous section we presented all the ingredients necessary to compute 
${\cal A}_{\Delta\Gamma}^{K^+ K^-}$ and ${\cal A}_{\Delta\Gamma}^{J/\psi f_0}$ as 
functions of $\phi_s$. Inserting them into the solution for $y_s$ discussed in 
Section~\ref{sec:cubic} allows us to draw contours for the lifetime measurements on the 
$\phi_s$--$\Delta\Gamma_s$ plane. The first measurement of the effective $B^0_s \to K^+ K^-$ 
lifetime was performed by the CDF collaboration in 2006 \cite{CDF-KK-lifetime}. In the spring 
of 2011, the LHCb collaboration reported their first measurement of this observable
\cite{LHCb-KK-lifetime}:
\begin{equation}\label{eqn:tauKK}
\tau_{K^+ K^-} = \left[1.44 \pm 0.096 {\rm (stat)}\pm 0.010 {\rm (syst)}\right] {\rm ps},
\end{equation}
which is currently the most precise.
The first measurement of the $B^0_s \to J/\psi f_0$ lifetime has recently been made by the
CDF collaboration \cite{Aaltonen:2011nk}:
\begin{equation}\label{eqn:tauf0}
\tau_{J/\psi f_0} = \left[1.70^{+0.12}_{-0.11} {\rm (stat)} \pm 0.03 {\rm (syst)}\right] {\rm ps}.
\end{equation}

\begin{figure}[tbp] 
   \centering
   	  \includegraphics[width=7.9truecm]{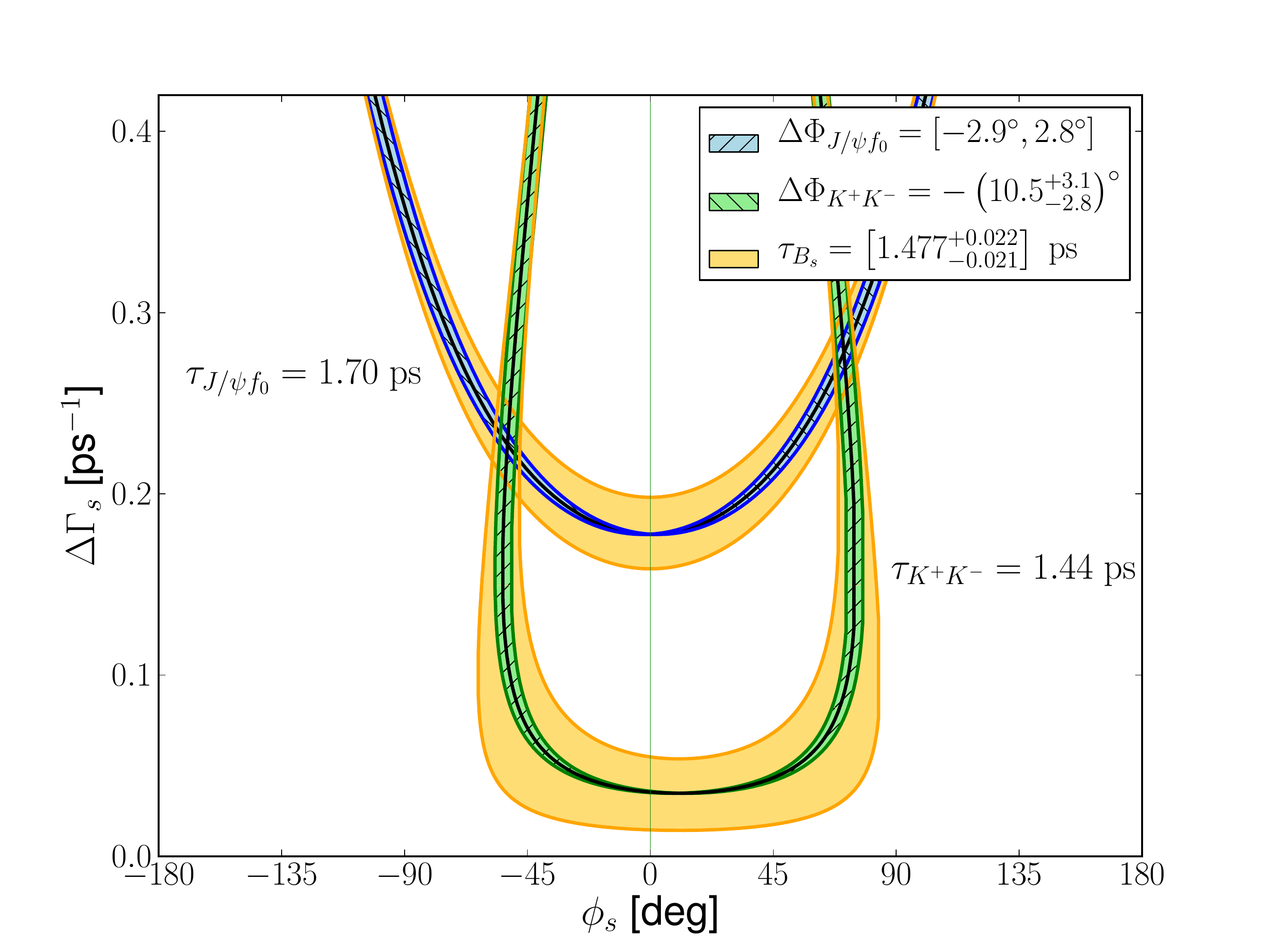} 
   \caption{Illustration of the errors of the hadronic phase shifts $\Delta\phi_{K^+K^-}$
   and $\Delta\phi_{J/\psi f_0}$ on the contours in the $\phi_s$--$\Delta\Gamma_s$ plane
   for the central values of the lifetime measurements. We also shown the impact of the 
   present error of the $B_s$ lifetime.}\label{fig:DGphiS-errors}
\end{figure}

In the left panel of Fig.~\ref{fig:DGphiS}, we show the current measurements of the
effective lifetimes of the $B^0_s \to K^+ K^-$ and $B^0_s \to J/\psi f_0$ decays as 
constraints on the $\phi_s$--$\Delta\Gamma_s$ plane. We also show the 39\% confidence 
region resulting from a $\chi^2$ fit of these two results. The individual fitted values for the 
$\phi_s$ and $\Delta\Gamma_s$ parameters are given as follows:
\begin{equation}\label{fit-1}
	\phi_s = -\left(52_{-43}^{+19}\right)^\circ,\quad 
	\Delta\Gamma_s = \left(0.23^{+0.08}_{-0.12}\right) {\rm ps}^{-1}
\end{equation}
\begin{equation}\label{fit-2}
	\phi_s = \left(71^{+14}_{-27}\right)^\circ,\quad
	\Delta\Gamma_s = \left(0.28^{+0.08}_{-0.14}\right) {\rm ps}^{-1},
\end{equation}
where the errors are 68\% confidence levels corresponding to a $\chi^2$ fit of the lifetimes.
Each solution has a two-fold ambiguity given by the transformation
\begin{equation}
	\phi_s \to \phi_s + 180^\circ,\quad \Delta\Gamma_s \to -\Delta\Gamma_s.
\end{equation}

Both lifetime measurements currently have an error of about 7\%. However, it seems
feasible to reduce the uncertainty of the $\tau_{K^+K^-}$ measurement at LHCb to the 
few-percent level \cite{Gersabeck}. In the right panel of Fig.~\ref{fig:DGphiS}, we 
show --  for illustration -- the impact of measurements of the $B^0_s\to K^+ K^-$ and 
$B^0_s\to J/\psi f_0$ lifetimes with $1\%$ uncertainty, assuming no change in the
central values. Clearly, at this level of accuracy, the lifetime measurements could 
strongly constrain $\phi_s$ and $\Delta\Gamma_s$. 

Using (\ref{ys}), we also include the band corresponding to the theoretical value of 
$\Delta\Gamma_s^{\rm SM}/\Gamma_s$ given in (\ref{eqn:DGGSM}). We observe,
as also noted in Ref.~\cite{FKR}, that the central value of the $\tau_{J/\psi f_0}$ measurement
is too large in comparison with this constraint. 
To spoil the relation in (\ref{ys}) either large NP effects are required, a very contrived 
scenario in our opinion, or the width difference $\Delta\Gamma_s$ must be affected by hadronic 
long-distance effects, which are not included in the SM calculation of (\ref{eqn:DGGSM}). 
The $B_s^0 \to J/\psi f_0$ effective lifetime predicted by the SM calculation is 
$\tau_{J/\psi f_0}=(1.582\pm0.036)$\,ps \cite{FKR}.

\begin{figure}[tbp] 
   \centering
	  \includegraphics[width=7.9truecm]{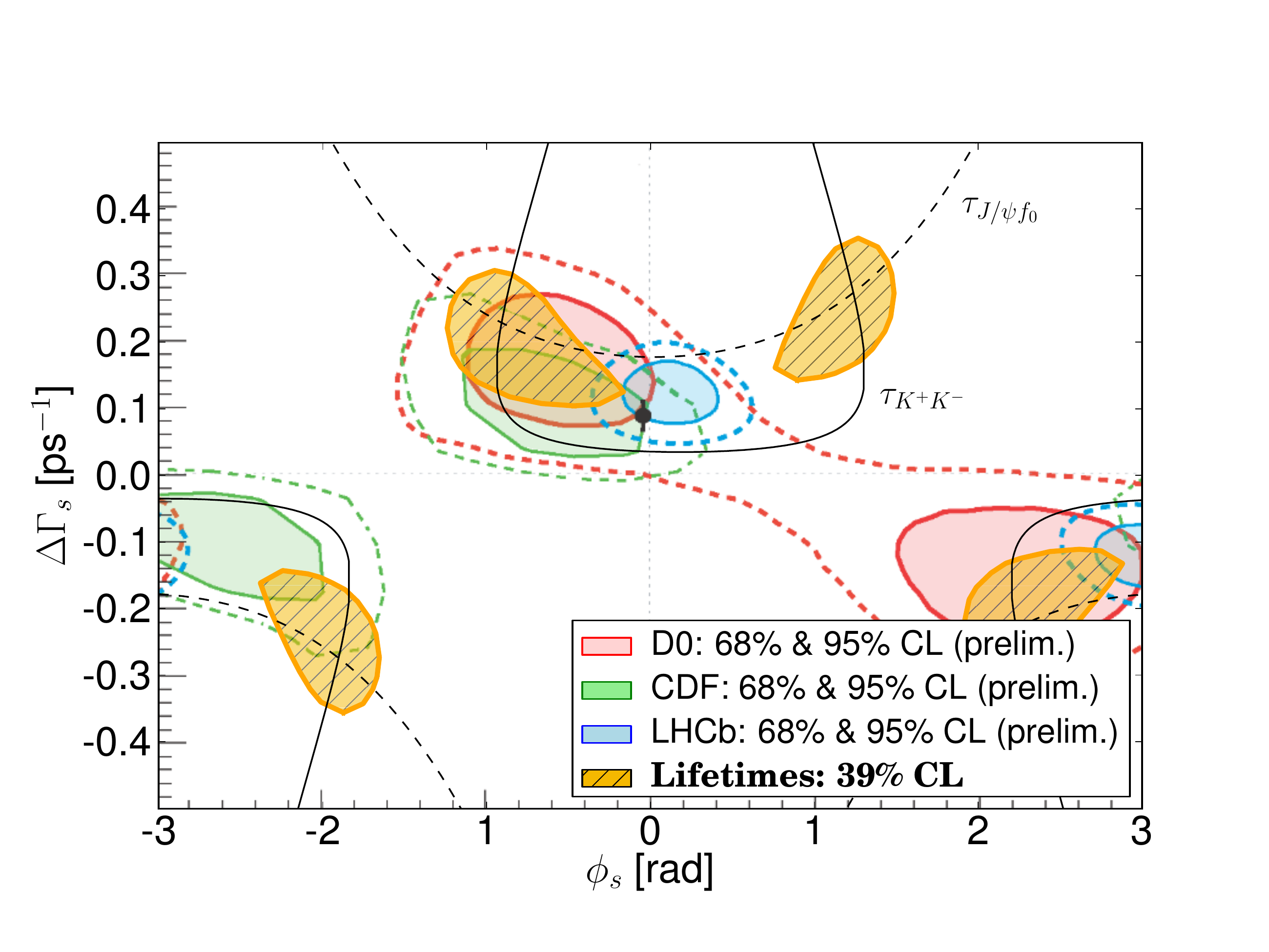} 
   \centering
   \caption{The fitted lifetime regions in the $\phi_s$--$\Delta\Gamma_s$ plane from the 
   left panel of Fig.~\ref{fig:DGphiS} added to a compilation of measurements as obtained 
   in Ref.~\cite{LHCb-public}. The \D0, CDF and LHCb allowed regions refer to tagged 
   analyses of $B^0_s\to J/\psi\phi$. In addition, the \D0 region includes also the result for 
   the like-sign dimuon asymmetry  while LHCb has also included a first 
   analysis of CP violation in $B^0_s\to J/\psi f_0$.}\label{fig:fits}
\end{figure}

The uncertainties of the hadronic phase shifts given in \eqref{eqn:DelPhif0} and 
\eqref{eqn:DelPhiKK} as well as the error of the $B_s$ lifetime in \eqref{eqn:tauBs} 
were not included in Fig.~\ref{fig:DGphiS} or in the fit results in (\ref{fit-1}) and (\ref{fit-2}). 
In Fig.~\ref{fig:DGphiS-errors}, we illustrate the impact of these uncertainties on the
lifetime contours in the $\phi_s$--$\Gamma_s$ plane. Comparing with the error bands
in Fig.~\ref{fig:DGphiS}, we observe that the effects of these uncertainties are marginal 
with respect to the current errors of the effective lifetime measurement. More sophisticated
fits should take these uncertainties into account as well. 

It is interesting to compare our fitted results to recent measurements of CP violation in the 
$B^0_s \to J/\psi \phi$ channel, where the current picture of $\phi_s$ looks as follows: 
CDF finds $\phi_s\in[-177.6^\circ,-123.8^\circ] \lor [-59.6^\circ,-2.3^\circ]$ (68\% C.L.), 
while \D0 has recently reported $\phi_s=-\left(31.5^{+20.6}_{-21.8}\right)^\circ$ 
\cite{TEV-LP}. The LHCb collaboration has also entered the arena, reporting 
$\phi_s=+(7.4\pm10.3\pm4.0)^\circ$ and 
$\Delta\Gamma_s=\left[0.123\pm0.029 {\rm (stat)}\pm0.008 {\rm (syst)} \right] {\rm ps}^{-1}$
\cite{LHCb-LP}. Furthermore, LHCb has presented a first tagged analysis of the
CP-violating asymmetry of the $B^0_s\to J/\psi f_0$ channel, yielding 
$\phi_s=-(25\pm25\pm1)^\circ$ \cite{LHCb-f0-conf}.

A compilation of the preliminary results from the \D0, CDF and LHCb collaborations as
constraints in the $\phi_s$--$\Delta\Gamma_s$ plane has recently been performed in 
Ref.~\cite{LHCb-public}. In Fig.~\ref{fig:fits}, we have overlaid on the corresponding plot 
the lifetime contours and fit results of the analysis described above. It is intriguing to see 
how well the lifetime allowed region overlaps with those from \D0 and CDF. 

The current errors leave space for interesting future developments. Should the central 
values of the CP-even and CP-odd lifetimes approach the theoretical SM point, the 
power to pinpoint $\phi_s$ is lost, as illustrated by Fig.~\ref{fig:CPillust}. However, 
because the curves are flat at this point, $\Delta\Gamma_s$ could still be determined 
accurately in this case.

\boldmath
\section{Further Promising $B_s$ Decays}\label{sec:future}
\unboldmath
So far we have discussed the effective lifetimes for the $B^0_s \to K^+ K^-$ 
and $B^0_s \to J/\psi f_0$ channels, which have both been measured
in first analyses by the CDF and LHCb collaborations. These channels have 
final states with opposite CP eigenvalues and happen to be well paired for 
obtaining constraints in the $\phi_s$--$\Delta\Gamma_s$ plane using the 
strategy proposed in this paper. The hadronic corrections in $B^0_s\to K^+K^-$
can be controlled even better in the future through precise measurements of the
CP-violating observables of the $B^0_d\to \pi^+\pi^-$ channel. Regarding
$B^0_s\to J/\psi f_0$, hadronic corrections have a minor impact on the lifetime
analysis. A potential control channel is $B^0_d\to J/\psi f_0$, although here the situation 
is much more involved than in $B^0_s\to K^+K^-$ due to the unsettled 
hadronic structure of the scalar $f_0(980)$ state \cite{FKR}. 

Another interesting decay that can soon be added to this picture is 
$B^0_s\to J/\psi K_{\rm S}$ \cite{RF-BsJpsiKS,DeBFK}, which has been observed 
by the CDF and LHCb collaborations \cite{CDF-BsJpsiKS,LHCb-BsJpsiKS}.
This channel has a final state with CP eigenvalue $-1$ and is caused by 
$\bar b\to \bar c c \bar d$ quark-level processes, i.e.\ it has a CKM structure that 
is different from the decays considered above. In particular, the relevant hadronic 
parameter does not enter in a doubly Cabibbo-suppressed way. However, the 
uncertainties can be controlled through $B^0_d\to J/\psi \pi^0$  and are found to have 
a moderate impact on the effective $B^0_s\to J/\psi K_{\rm S}$ lifetime \cite{DeBFK-CKM}, 
which has not yet been measured.

Another $B_s$ decay with a CP-even final state is $B^0_s\to D_s^+D_s^-$. Here the
hadronic corrections are again doubly Cabibbo-suppressed and can be controlled
with the help of the $U$-spin-related $B^0_d\to D^+D^-$ decay \cite{RF-BsJpsiKS}.
A first theoretical analysis of the effective lifetime of $B^0_s\to D_s^+D_s^-$ was performed
in Ref.~\cite{RF-BsDD}. Further decays into CP-even final states where a similar analysis 
can be performed are the $B^0_{s(d)}\to J/\psi \eta^{(')}$ channels \cite{skands}.

Decays of $B_s$ mesons into CP-selfconjugate final states with two vector mesons 
or higher resonances offer another laboratory for lifetime analyses. In this case the 
$B^0_s\to \phi\phi$ and $B^0_s\to K^{*0}\bar K^{*0}$  channels look particularly interesting.
These decays, which have already been observed experimentally  
\cite{TEV-LP,LHCb-LP,LHCb-Kast0Kast0}, are penguin modes. Their final 
states are mixtures of CP-even and CP-odd eigenstates and can be disentangled by 
means of angular analyses. It would be interesting to perform measurements of the 
lifetimes for the CP-even and CP-odd final-state configurations and to add them as 
contours to the $\phi_s$--$\Gamma_s$ plane along the lines of the strategy proposed 
above. 

A similar comment applies to the $B_s \to J/\psi \phi$ channel, where it 
would also be desirable to determine the individual lifetimes for the CP-even 
and CP-odd final-state configurations separately instead of making a fit to the 
whole time-dependent angular distribution. This can be done by means of the 
moment analysis proposed in Ref.~\cite{DDF}. The hadronic uncertainties of the 
$B^0_s \to J/\psi \phi$ channel can be controlled by channels such as 
$B^0_s\to J/\psi \bar K^{*0}$ and $B_d^0 \to J/\psi \rho^0$ \cite{FFM}.

In the future, also decays with final states that are not CP-selfconjugate can be added 
to the agenda to further constrain $\phi_s$ and $\Delta\Gamma_s$, provided both a 
$B^0_s$ and a $\bar B^0_s$ meson can decay into the same final state (see 
Section~\ref{sec:effective}). Prime examples are the $B_s\to D_s^\pm K^{(*)\mp}$ channels. 
Their effective lifetimes can be used to constrain $\phi_s + \gamma$ with respect to 
$\Delta\Gamma_s$ (see Ref.~\cite{RF-BsDsK} for an overview of the observables of these 
decays).

\section{Conclusions}\label{sec:concl}
Thanks to the sizable width difference $\Delta\Gamma_s$ of the $B_s$-meson system,
effective lifetimes of $B_s$ decays offer interesting probes of $B^0_s$--$\bar B^0_s$ mixing.
The corresponding measurements require only untagged data samples and are advantageous
from an experimental point of view. Thanks to non-linear terms in $\Delta\Gamma_s$, a
pair of $B_s$ decays into CP-even and CP-odd final states is sufficient to determine the 
$B^0_s$--$\bar B^0_s$ mixing phase and the width difference. Prime examples for implementing
this strategy in practice are the decays $B^0_s\to K^+K^-$ and $B^0_s\to J/\psi f_0$. Their 
effective lifetimes turn out to be robust with respect to hadronic uncertainties, which can 
be controlled or constrained with the help of further experimental data. 

We have calculated the constraints in the $\phi_s$--$\Delta\Gamma_s$ plane following from
the current measurements of the $B^0_s\to K^+K^-$ and $B^0_s\to J/\psi f_0$ lifetimes, which both
suffer from $\sim~7\%$  uncertainties. The resulting picture is consistent with other constraints
following from Tevatron measurements, which have not been supported by recent LHCb data.
The uncertainties still preclude us from drawing definite conclusions but leave space 
for interesting future developments. Lifetime measurements with $1\%$ precision
would allow us to obtain much stronger constraints in the $\phi_s$--$\Delta\Gamma_s$ plane 
as we have illustrated in our study. Should the central values of the lifetimes for the CP-even
and CP-odd final states approach the SM values, the lifetime contours would loose their
power to determine $\phi_s$. However, the width difference could still be determined in a 
precise way in this case. 

An interesting trend of the current data is that it favours a value of $\Delta\Gamma_s$ 
that is larger than the one calculated in QCD. With the plausible assumption that NP affects this 
observable only through $B^0_s$--$\bar B^0_s$ mixing, its absolute value can only decrease
and hence the discrepancy will be even larger for sizable mixing phases. 
This feature raises the question of whether the SM calculation of $\Delta\Gamma_s$
fully includes all hadronic long-distance contributions. It will be interesting to see if 
this trend will be supported by future data or if it will eventually disappear. 

In the future, also other effective lifetime measurements of $B_s$-meson decays 
can be added to the $\phi_s$--$\Delta\Gamma_s$ plane, allowing us to ``overconstrain" 
the mixing parameters in the same spirit as the determination of the apex of the unitarity triangle. 
This information will be complementary to the tagged analyses of CP violation in the $B_s$ system. 
It will be intriguing to see at which point of the $\phi_s$--$\Delta\Gamma_s$ plane all 
measurements will eventually converge.


%
%
%
\end{document}